**Article type: Progress Report**

## Two-dimensional layered materials for memristive and neuromorphic applications

*Chen-Yu Wang[#], Cong Wang[#], Fanhao Meng[#], Pengfei Wang, Shuang Wang, Shi-Jun Liang\*, Feng Miao\**

C-Y. Wang, C. Wang, F. Meng, P. Wang, S. Wang, Dr. S-J. Liang, Prof. F. Miao
National Laboratory of Solid State Microstructures, School of Physics, Collaborative Innovation Center of Advanced Microstructures, Nanjing University, Nanjing 210093, China

Corresponding author: sjliang@nju.edu.cn; miao@nju.edu.cn
[#]The authors C-Y. Wang, C. Wang and F. Meng made equal contribution to this work.

**Abstract**
With many fantastic properties, memristive devices have been proposed as top candidate for next-generation memory and neuromorphic computing chips. Significant research progresses have been made in improving performance of individual memristive devices and in demonstrating functional applications based on small-scale memristive crossbar arrays. However, practical deployment of large-scale traditional metal oxides based memristive crossbar array has been challenging due to several issues, such as high-power consumption, poor device reliability, low integration density and so on. To solve these issues, new materials that possess superior properties are required. Two-dimensional (2D) layered materials exhibit many unique physical properties and show great promise in solving these challenges, further providing new opportunities to implement practical applications in neuromorphic computing. Here, recent research progress on 2D layered materials based memristive device applications is reviewed. We provide an overview of the progresses and challenges on how 2D layered materials are used to solve the issues of conventional memristive devices and to realize more complex functionalities in neuromorphic computing. Besides, we also provide an outlook on exploitation of unique properties of 2D layered materials and van der Waals heterostructures for developing new types of memristive devices and artificial neural mircrocircuits.

## 1. Introduction

Memristive device, also known as memristor and ReRAM, was originally conceived in 1971[1] and then experimentally realized in 2008[2]. The sandwich structure of memristive devices is usually composed of an insulating active layer (*e.g.*

transition metal oxides (TMOs)) and top/bottom metal electrodes. The electrical resistance of the devices relies on the history of current or voltage previously applied. It has been demonstrated that the memristive devices exhibit many promising features [3-8], such as non-volatility, high speed switching, high endurance, high-density integration, CMOS-compatible fabrication and so on. These features make them ideal candidates for applications in memory devices [3, 5, 9], in-memory computing [3, 10-12], and edge computing[13, 14]. The working principle of the TMOs-based memristive devices relies on ion drift or diffusion, which resembles motion of ions in the biological neurons and synapses. The ionic motions exhibit dynamic behaviors. Thus, the memristive devices can be used to emulate the synaptic plasticity[15] and neurons[16-18]. Compared to traditional silicon synapses [19-21] and neurons[22, 23] requiring complex circuits, such emerging memristive devices have compact structures and enhanced functionalities in mimicking synaptic plasticity and neurons, and have been widely applied in the realization of artificial neural network (ANN) accelerator. [3, 12, 24-27] Assembling the memristive devices into arrays enables to perform the operation of dot multiplication based on Kirchhoff's law, which is the most complex operation in ANN. Compared to that implemented in von Neumann architecture, this multiplication can be accelerated in hardware by feeding voltage signals representing vectors into crossbar array representing matrix. This approach has been demonstrated in sparse coding[28], convolution neural network (CNN)[14], spike neural network (SNN)[29], reinforcement learning[30]. Although the memristive devices hold promise in many applications, it has been challenging to improve the performance of TMOs based devices [11] in aspects of power consumption, device reliability and transparency/flexibility. For example, many TMOs-based memristive devices show low ON/OFF resistance, leading to an increase in the operating power during SET/RESET processes. The issue of high-power consumption becomes more serious in ANN accelerator as many SET/RESET processes are involved in neural network training. In addition to the power consumption, poor device reliability also restricts the practical deployment of large-scale array devices, which arises from random growth of conductive filaments (CFs) in active layer under action of vertical electric field[31]. To overcome these challenges, new materials and new device structures, as well as new working principles, are required.

Two-dimensional (2D) layer-structured materials newly emerge and exhibit a large diversity of physical properties. Gate-tunable physical properties enable these atomically-thin layered materials[32] to be promising in the design of diverse electronic and optoelectronic devices[33-41]. The diversity can be further enhanced by assembling distinct 2D materials to van der Waals (vdW) heterostructures[41-45]. The unique properties that 2D materials and vdW heterostructures exhibit may be utilized to solve the aforementioned issues faced by conventional memristive devices[46-52]. For example, graphene is semimetallic and suitable for electrodes of the memristive devices. The low density of state (DOS) in the graphene effectively suppresses operating current flowing through the devices and reduces power consumption[53]. Due to high electron-density induced by $sp^2$ hybridized carbon atoms and strong in-plane covalent atomic structure, almost all molecules and ions are impermeable across graphene[54]. With this intrinsic property, ionic griddle for improving devices reliability can be fabricated through defects engineering to control the shape and location of conductive filaments[55]. Compared to conventional TMOs materials, 2D layer-structured materials show much better mechanical strength[32, 56]. The excellent mechanical property makes memristive devices based on pure 2D vdW heterostructure exhibit stable resistive switching behaviors against mechanical stress[57], indicating huge potential in flexible electronics applications. Although the growth of conductive filaments and electrochemical dynamics of nanoscale clusters in the TMOs-based memristive devices have been studied by transmission electron microscopy,[7, 31, 58, 59] further understanding of switching mechanism may benefit from the use of transparent 2D materials. With graphene as transparent electrode, conductive atomic force microscope (CAFM)[60] and Raman spectrum [61] can be used to locate the filaments and their morphology. *In operando* spectromicroscopy[62] has been used to reveal the redox reaction driven switching. Recently, electrical field induced phase change in the lattice of 2D materials has been verified by high-angle annular dark field (HAADF) STEM[57, 63]. These characterizations are crucially important to deeply understand the working mechanisms of memristive and neuromorphic devices.

Although 2D layered materials hold promise in solving challenges faced by TMOs-based memristive devices and synaptic devices, there are still a few key challenges to overcome. The unique physical properties that 2D layered materials and vdW heterostructure possess have not been fully exploited for designing neuromorphic devices with new working principles. Memristive devices relying on charge-trapping effect[64-70] at the interface are not suitable for development of robust devices, as it is difficult to control the trapping/detrapping process in a reliable way. Therefore, it is desirable to develop novel memristive devices based on distinct working principles. Besides, the studies of neuromorphic devices have been mainly focusing on individual synaptic devices. Fabrication of large-scale device arrays has seldom been demonstrated. This is associated with the challenge in synthesis of large area 2D materials and vdW heterostructures. Although synthesis of large-area monolayer graphene or h-BN has been already demonstrated, it is still a big challenge to assemble various 2D materials to vdW heterostuctures. Note that although solution-processed 2D materials[51, 71-74] show certain promise in fabricating large-scale memristive and synaptic devices, we will focus our attention in the field of memristive devices based on 2D layered materials in this report. Those progresses along solution-processed 2D materials will not be reviewed here. Interested readers may refer to the previous reviews [75, 76] for more details. Moreover, the applications of memristive devices based on 2D layered materials or vdW heterostructures are mainly limited to emulation of biological synaptic behaviors, with few works emulating behaviors of biologic neurons. This would lead to a challenge in constructing neural circuits, which are regarded as building blocks of information processing mimicking the neural network of human brain.

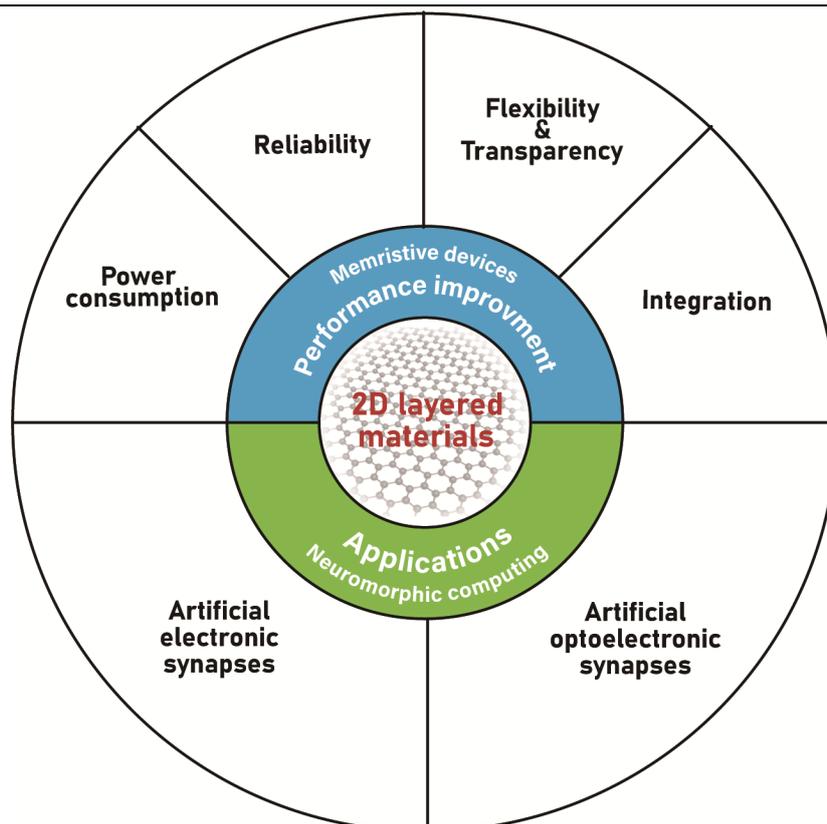

Figure 1. A summary of two-dimensional layered materials used for memristive and neuromorphic applications

In this review, we provide an overview of how 2D layered materials can be used to overcome the issues faced with TMOs-based memristive devices and to develop novel neuromorphic devices of distinct working principles, which are graphically summarized in Fig. 1. In Section 2, we will review the research progress on the use of 2D layered materials in improving device performance, *e.g.* power consumption, reliability, flexibility, transparency as well as integration. In Section 3, the recent studies of 2D materials based synaptic devices with new structures and new mechanisms will be reviewed. Finally, we will discuss the challenges and opportunities associated with 2D materials based memristive devices and their applications in neuromorphic computing.

## 2. Memristive devices

With the advance of intelligent terminals such as cell phones and wearable devices, the concept of edge computing has been proposed to solve the challenge of mass data processing by endowing edge devices with ability to process data[13, 77, 78]. The edge computing could alleviate burdens of data transmission between edge devices and large data center and accelerate data processing in large data centers. Due to the potential in processing in memory, memristive devices crossbar arrays are suitable for high-efficiency information processing without shuttling data back and forth[3] between memory and processing units, making them promising candidates in edge computing[13, 14]. Nevertheless, the challenges in power consumption, device reliability and high-density integration hinder the development of TMOs based memristive devices. Compared to TMOs, 2D layered materials show more diverse physical properties and have huge potential in overcoming these challenges.

## 2.1. Power consumption

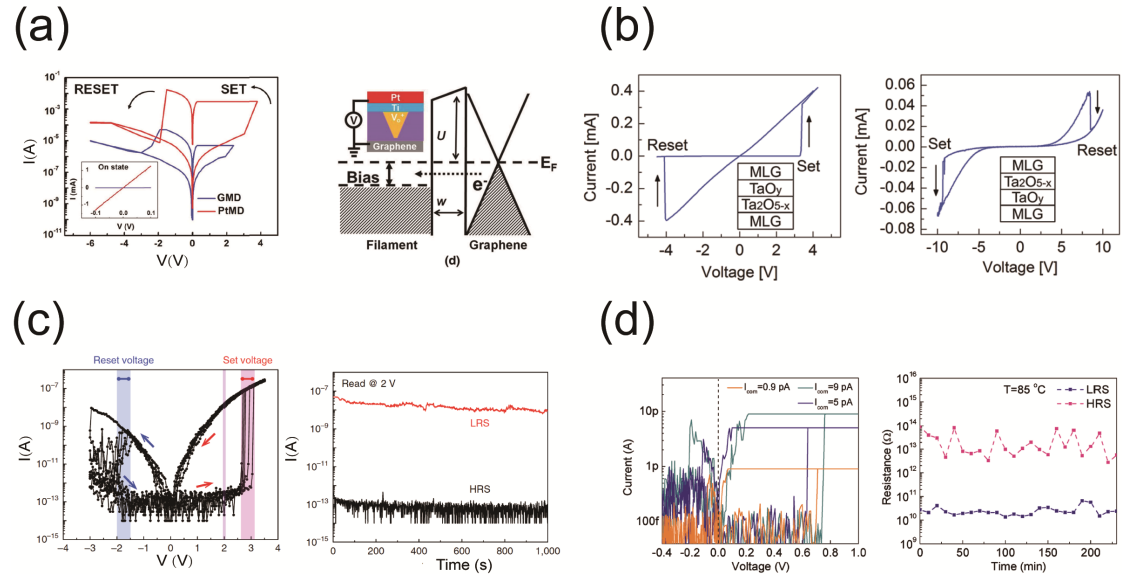

Figure 2. 2D layered materials for lowering power consumption of memristive devices. (a). Memristive devices based on Gr/TiO$_x$/Ti/Pt. Left panel: switching curves of graphene-based memristive devices (GMDs) and Pt-based memristive devices (PtMDs); right panel: illustration of GMDs band structure at positive bias. (b). Memristive devices based on Gr/TaO$_y$/Ta$_2$O$_{5-x}$ /Gr devices and Gr/Ta$_2$O$_{5-x}$/TaO$_y$/Gr devices, and their switching curves. (c). Memristive devices based on bilayer compose of MoTe$_2$ flake and Al$_2$O$_3$ film. Left panel: switching curves; right panel: retention time of ON/OFF states. (d). Memristive devices based on h-BNO$_x$. Left panel: switching curves at different compliance currents; right panel: retention test at 85 °C. (a), reproduced with permission[53]. Copyright 2014, John Wiley & Sons, Inc. (b), reproduced with permission[79]. Copyright 2014, John Wiley & Sons, Inc. (c), reproduced with permission[63]. Copyright 2019, Springer Nature Publishing AG. (d), reproduced with permission[80]. Copyright 2017, John Wiley & Sons, Inc.

High power consumption of many TMOs based memristive devices arises from low ON/OFF resistance state, which could be effectively reduced by decreasing current flowing through memristive devices. Graphene has a linear band structure and a low DOS[81]. Using graphene to replace tungsten as electrode results in a low switching current and reduced power consumption in Dy$_2$O$_3$ and Gr/TiO$_x$/Al$_2$O$_3$/TiO$_2$/Gr[82] based memristive devices[83]. Qian et al. have achieved a low switching current (1 μA) [53] in Gr/TiO$_x$/Ti/Pt devices (Figure 2(a)). The reduction in switching current is attributed to the low DOS near Dirac cone of graphene. This low DOS gives rise to a small electron tunneling probability from metallic filaments to graphene. As a result, the SET current is reduced by 600 times compared to that of Pt-based memristive devices. The operating current could be further reduced by 50 times when lower quality graphene is used, which may derive from a reduction of the carrier tunneling probability. In addition to suppressing the switching current, the device shows strong nonlinear current-voltage (I-V) characteristics. The ON-state resistance is about $10^4 \Omega$, much larger than that of Pt-based counterparts ($10^2 \Omega$). This property can be used to enhance nonlinearity of selectors and solve the sneak path issue in large-scale array by connecting with memristive devices in series. By connecting an $Au/TiO_x/Gr/TiO_x/Au$ based selector with a $TaO_x$ based memristive device into a one-selector one-memristor (1S1R) cell, Wang et al. demonstrated nonlinear and symmetric I-V switching characteristics[84]. Apart from graphene, insulating graphene oxides (GOs)[72] could also be used to reduce the operating current and enhance the nonlinearity of I-V characteristics. The electrical behaviors of $TaO_y/Ta_2O_{5-x}$ bilayer devices with graphene electrodes vary on the deposition sequence of $TaO_y$ and $Ta_2O_{5-x}$ films (Figure 2(b))[79]. During the fabrication of $Gr/TaO_y/Ta_2O_{5-x}/Gr$ devices, pre-deposition of $Ta_2O_{5-x}$ in inert argon (Ar) plasma results in classical bipolar resistive switching, similar to the devices with metal electrodes. However, devices show distinct I-V characteristics when $TaO_y$ film was first deposited. The reason is that exposing graphene bottom electrode (GBE) to oxygen (O) plasma environment for the growth of $TaO_y$ film would cause partial oxidization of graphene. Charge tunneling through the partially oxidized graphene electrode dominates the I-V characteristics of memristive devices. Therefore, the $Gr/Ta_2O_{5-x}/TaO_y/Gr$ devices can be regarded as a highly compact 1S1R cell with strong I-V nonlinearity.

In addition to electrode material, 2D materials could also be used as active layer to increase resistance of memristive devices[63] and reduce operating current. By inserting a tunneling barrier layer (i.e. $Al_2O_3$ film) between $MoTe_2$ (6~36 nm) and electrode to suppress the tunneling current, Zhang et al. fabricated a novel memristive device. This device shows excellent performance, such as an ON/OFF ratio of $10^5 \sim 10^6$ (Figure 2(c)), large nonlinearity at low resistance state (LRS), and high resistance state (HRS) resistance of larger than 10 T$\Omega$. Different from previous work that electrostatic doping in the monolayer $MoTe_2$ drives a phase transition from 2H (semiconducting) to 1T' (metallic)[85], high-angle annular dark field (HAADF) STEM characterizations indicate that that resistive switching in thick $MoTe_2$ based memristive device can be attributed to electric-field-induced lattice phase transition between $2H_d$ and 2H. These findings demonstrate that controlled phase transition in 2D materials could also be used as switching mechanism of memristive devices. Without tunneling barrier layer, the memristive devices that use insulating 2D materials, like h-BN[80] and $(PEA)_2PbBr_4$[86], as active layer also show low power consumption. For instance, h-BN could be oxidized to h-$BNO_x$ by ultralow-power oxygen plasma to suppress the tunneling current in vertical direction[87, 88]. The Ag/h-$BNO_x$/Gr devices with different thickness of h-$BNO_x$ flakes (0.9~2.3 nm) show reliable switching behaviors and ultralow SET/RESET current (Figure 2(d)). The operating current can be reduced down to sub-pA level in the devices with atomically thin active layer. The power consumption is reduced to a range of 100 aJ to 1 fJ, which is the lowest record reported for memristive devices. Therefore, it is very promising to apply 2D material based memristive devices with low power consumption to practical application, such as aerosolizable electronics [89].

## 2.2. Device reliability

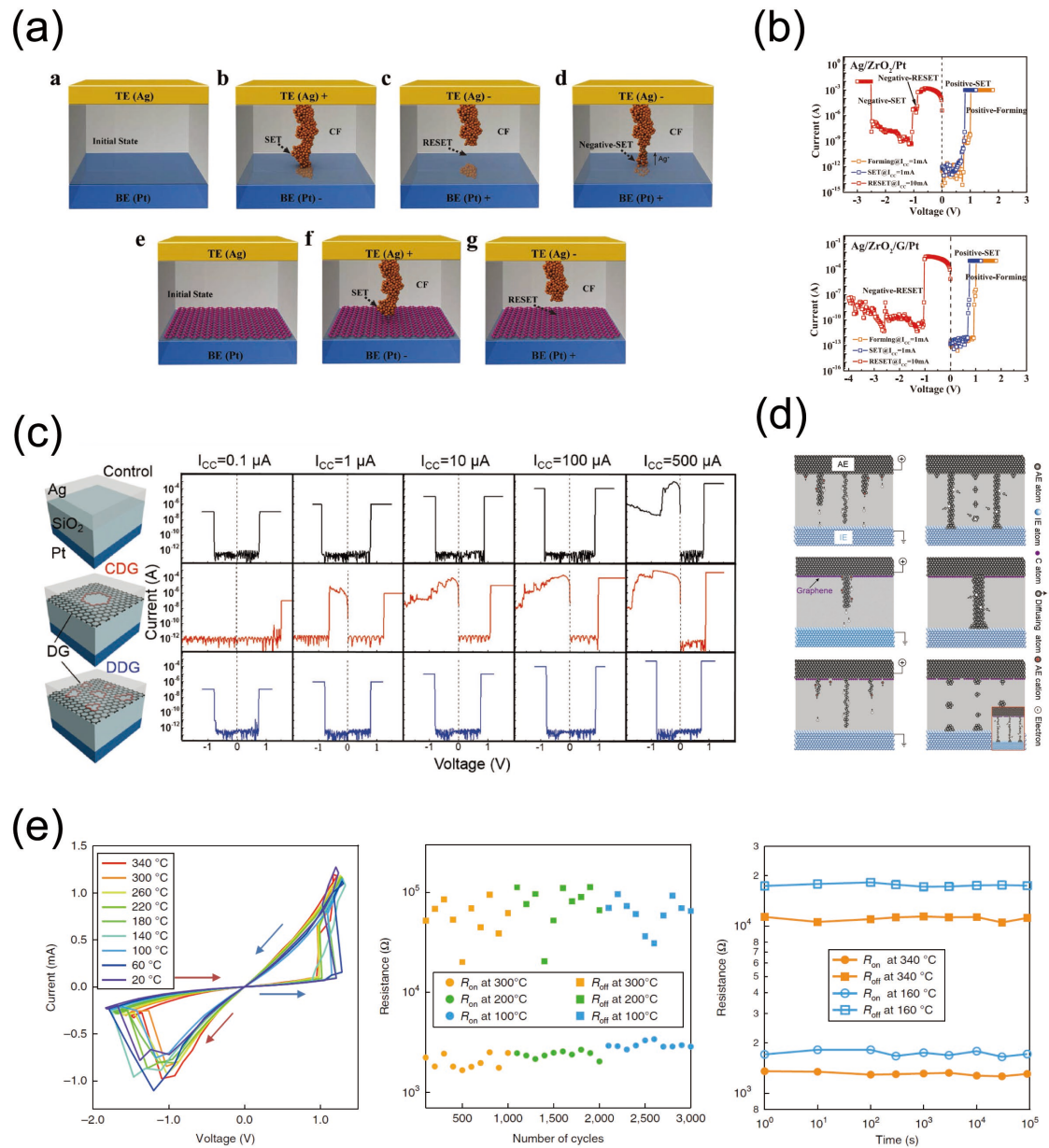

Figure 3. 2D layered materials for improving device reliability. (a). Illustration of the work principle of Ag/ZrO₂/Pt and Ag/ZrO₂/Gr/Pt devices. Use of graphene prevents the growth of filament near bottom electrode to eliminate negative-SET. (b). Comparison of switching curves between Ag/ZrO₂/Pt and Ag/ZrO₂/Gr/Pt devices. (c). Switching curves of Ag/SiO₂/Gr/Pt (concentrated-defected and discrete-defected graphene respectively) and Ag/SiO₂/Pt devices (control devices) under different current compliance. (d). Illustrations of filament formation and rupture in three types of devices mentioned in (c). (e). Performance of memristive devices based on Gr/MoS₂₋ₓOₓ/Gr vdW heterostructures at different temperatures. (a)-(b), reproduced with permission[90]. Copyright 2016, John Wiley & Sons, Inc. (c)-(d), reproduced with permission[55]. Copyright 2018, John Wiley & Sons, Inc. (e), reproduced with permission[57]. Copyright 2018, Springer Nature Publishing, AG.

The controllable formation and rupture of conductive filaments is critical for the development of high-performance memristive devices[25]. The overgrowth of filaments usually drives excessive ions into inert electrode[91-93] and leads to negative-SET behavior in cation-based memristive devices. The uncontrollable migration of ions in the active layer under an electric field[31, 55, 90] would lead to a current-retention dilemma in memristive devices. Graphene has a hexagonal lattice structure with an intrinsic pore size of 0.064 nm[94-96], and the formation energy of defects in the graphene lattice is very high (about 8.0 eV)[90, 95]. Thus, graphene is an ideal blocking layer to ions diffusion. Liu et al. have shown that electrical field driven metal ions in active layer cannot penetrate through graphene to reach electrodes[90]. At positive bias, excessive active metal atoms enter into the inert electrodes. As a result, redox reaction of active metal atoms at negative bias takes place in the inert electrode and leads to the formation of undesired filaments as shown in Figure 3(a). Inserting graphene as blocking layer between active layer and electrode allows for the control of metal ions migration in active layer and may eliminate the negative-SET behaviors (Figure 3(b)), which suggests that the use of graphene indeed prevents overgrowth of conductive filaments in active layer into inert electrode.

Current-retention dilemma occurring in memristive devices is another critical issue that needs be addressed to improve device reliability. Large operating current and long retention often coexist in memristive devices, which increase the difficulty to fabricate high-speed selectors with large operating current. Prior studies have shown that excellent impermeability of graphene can be used in improving the resistive switching properties[95]. By using two defective graphene layers, Zhao et al. have fabricated Pt/Ag/Gr/SiO$_2$/Pt resistive switching devices[55] (Figure 3(c)) to break the dilemma. To this end, two types of defects have been introduced in the graphene. The first one is nanoscale concentrated defect. It was created by electron beam lithography and oxygen plasma etching. In the devices with this type of defective graphene, the single conductive filaments only form in close proximity to the nano-holes. The stability of the conductive filaments is greatly enhanced due to small total surface area. On the contrary, the discrete tiny conductive filaments are formed in the devices with discrete-defect graphene. They are readily disrupted into discontinuous nanoparticles (Figure 3(d)) in a spontaneous way. By controlling the morphology and location of conductive filament formation, Zhao et al. finally implemented fast selectors with a high On-state current ($\approx$1 mA) and high selectivity ($5 \times 10^8$) as well as high switching speed (0.1/1 µs) in the Pt/Ag/Gr/SiO$_2$/Pt resistive switching devices.

Traditional TMOs-based memristive devices fail to operate at high-temperature conditions. This challenge can be overcome by using 2D layered materials, as 2D layer-structured materials retain excellent crystal structure at high temperature. Wang et al. demonstrated that memristive devices based on Gr/MoS$_{2-x}$O$_x$ (20~40 nm)/Gr (GMG) vdW heterostructures exhibit stable resistive switching. More interestingly, resistive switching is stable at high temperatures up to 340 ºC (Figure 3(e)), which is a record high among all memristive devices[57] reported so far. Furthermore, Wang et al. pointed out that such high-temperature robustness of vdW heterostructure memristive devices arises from the maintained excellent crystal structure of MoS$_{2-x}$O$_x$ at elevated

WILEY-VCH

temperatures[97, 98]. They also carried out *in-situ* STEM (scanning transmission electron miscroscope) studies to study the switching mechanism of the GMG devices. The experimental results indicate that sulfur and oxygen ions migration play important roles in the resistance change of $MoS_{2-x}O_x$ membrane. In pristine state, sulfur vacancies are occupied by oxygen ions and the averaged atomic ratio of molybdenum to sulfur and oxygen is around 1:2. After the electroforming process, the percentage of sulfur atom in the channel region is apparently reduced, as confirmed by EDS (energy dispersive spectroscopy) line-scan profiles. The ratio of molybdenum to sulfur and oxygen restores to 1:2 after RESET process, suggesting that oxygen atoms migrate into the channel region under the action of electrical field and fill sulfur vacancies in the RESET process. Since diffusion of oxygen ions has a lower energy barrier than that of sulfur ions, oxygen ions could be driven out of the channel region by the temperature gradient due to the dominated thermophoresis effect in SET process.

## 2.3. Flexibility and transparency

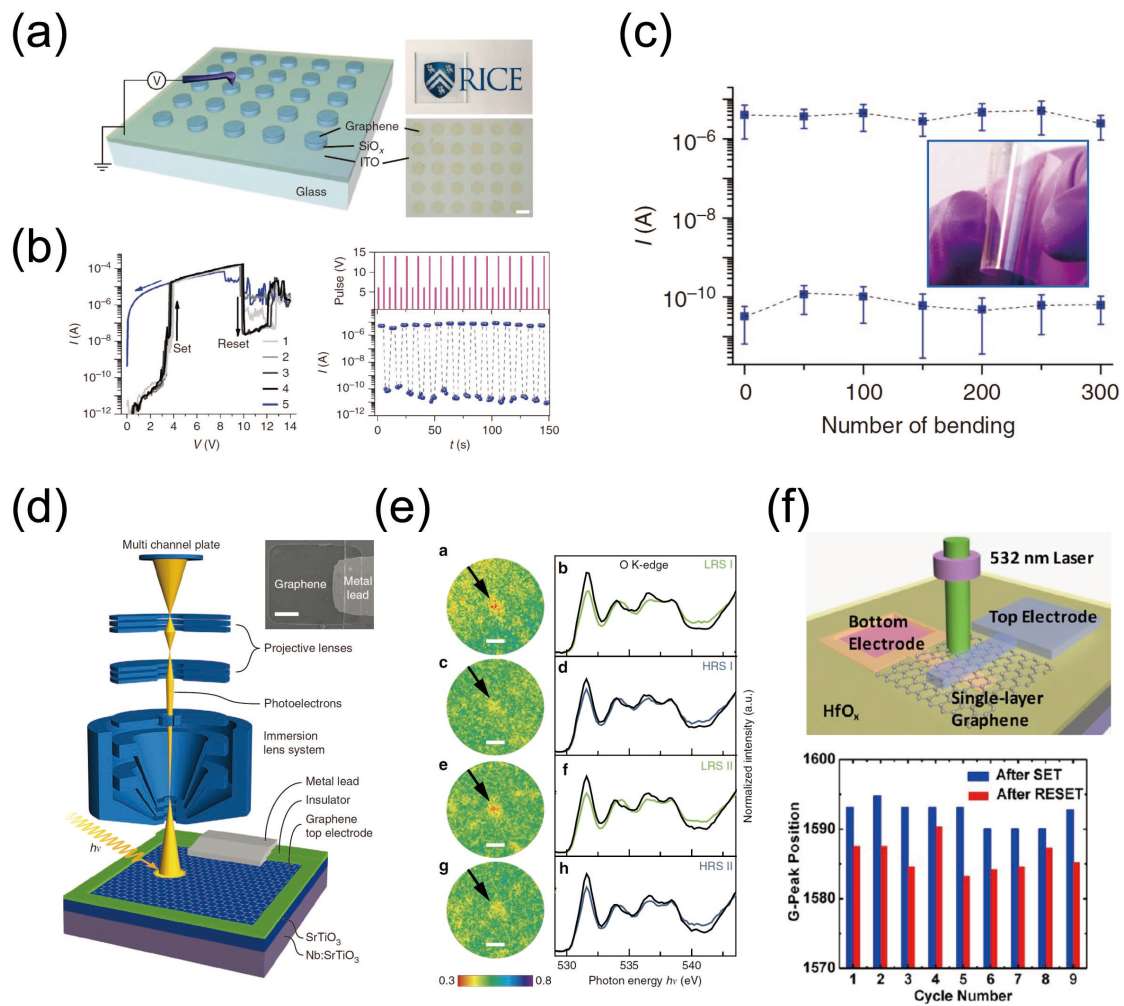

Figure 4. Flexibility and transparency of 2D layered materials for memristive devices. (a). Illustration and optical image of transparent $Gr/SiO_x/ITO$ devices on a glass substrate. Scale bar, 100 μm. (b). Left panel: unipolar switching curves of a $Gr/SiO_x/ITO$ device; right panel: current variation corresponding to a series of voltage pulses of +6 V, +1 V and +14 V. (c). Retention test of ON and OFF states of a Gr/SiO$_x$/Gr device against different bending cycles. (d). Set-up of *in operando* spectromicroscopy for studying memristive devices with graphene top electrodes. The inset is the SEM image of the device (scale bar, 5 μm). (e). Spectromicroscopic quantification of resistive switching filaments. (a,c,e,g): photoelectron emission microscopy (PEEM) images of switching filaments in LRS, HRS, LRS II and HRS II. Scale bars, 1μm; (b,d,f,h): O K-edge for switching filaments in (a,c,e,g). (f). Monitoring oxygen movement of devices with a graphene-inserted electrode by Raman spectroscopy. (a)-(c), reproduced with permission[99]. Copyright 2012, Springer Nature Publishing AG. (d)-(e), reproduced with permission[62]. Copyright 2016, Springer Nature Publishing AG. (f), reproduced with permission[61]. Copyright 2013, American Chemical Society Publishing.

Flexibility and transparency provide memristive devices with new opportunities in applications of wearable[100, 101] and transparent[102] electronics. 2D layered materials are known for their excellent mechanical properties[103, 104] and high optical transmittance[105], making them suitable for wearable and transparent electronic applications[71, 106, 107]. Compared to organic counterparts[108], 2D layered materials based memristive devices show better flexibility and transparency as well as performance (*e.g.* large ON/OFF ratio, long retention and high endurance, etc.). Using graphene and Indium tin oxide (ITO) as electrodes, SiO$_x$ based memristive devices with unipolar switching behavior possess a high optical transmittance at 550 nm up to 90%[99] (Figure 4(a)-(b)). Flexible and transparent Gr/SiO$_x$/Gr memristive devices fabricated on fluoropolymer substrate show stable switching behavior after 300 bending circles (Figure 4(c)). Compared to Gr/SiO$_x$/Gr memristive devices, Gr/Dy$_2$O$_3$/Gr memristive devices show higher ON/OFF ratio ($10^4$), faster switching speed (60 ns) and lower power consumption (100 μW)[83]. For oxide based memristive flexible devices, it is difficult to keep stable ON/OFF states after exceeding 1000 bending circles, while it is readily retainable in the vdW heterostructure based memristive devices[109]. Memristive devices based on GMG vdW heterostructures[57] exhibit almost identical switching curves after 1200 bending cycles. Unlike the TMOs-based or organic memristive devices, which can only retain robustness either in flexibility or thermal stability, vdW heterostructure based memristive devices simultaneously possess both flexibility and thermal stability and show promising applications in high-temperature electronics and various scenarios against mechanical stress.

Using transparent 2D materials as electrodes could also help reveal the detail of filament growth and rupture. High surface sensitivity of metal electrodes is the main obstacle for characterizing memristive devices with *in operando* spectromicroscopy. This challenge could be resolved by using graphene electrodes with high optical transmittance. Baeumer et al. have used *in operando* spectromicroscopy technology to study nanoscale redox reactions during resistive switching in G/SrTiO$_3$/Nb:SrTiO$_3$ structure (Figure 4(d))[62]. As shown in Figure 4(e), the change of O K-edge between ON and OFF states indicates that resistance change is dominated by redox reactions. The experimental results are consistent with simulation results in a quantitative manner, suggesting that changing donor concentrations at electrode-oxide interfaces could modulate effective Schottky barrier. Varying donor concentration by a factor of 2-3 leads to a change in device resistance by more than two orders of magnitude.

Transparency of graphene could also be used for observing $O^{2-}$ migration during switching. Large surface area enables graphene to serve as atomic and ionic reservoir, which results in a transition from valence change memories (VCM) to electrochemical metallization memories (ECM)[110]. It has been shown that graphene can be functionalized with epoxide groups to form modest and reversible covalent bonds. By inserting graphene at the electrode/oxide interface or using graphene as electrodes of memristive devices, researchers are able to use Raman spectroscopy to monitor oxygen movement *in-situ* through either area mapping or single-point measurements (top panel of Figure 4(f)) [61]. As shown in the bottom panel of Figure 4(f), G-peak position in Raman spectrum of monolayer graphene shifts after each SET/RESET process, which might be attributed to the reversible formation of covalent bonds at graphene surface. This approach provides an efficient and powerful technique to understand the process of oxygen movement at the interface, and has been widely used in studying other graphene based memristive devices[82, 111].

## 2.4. Integration

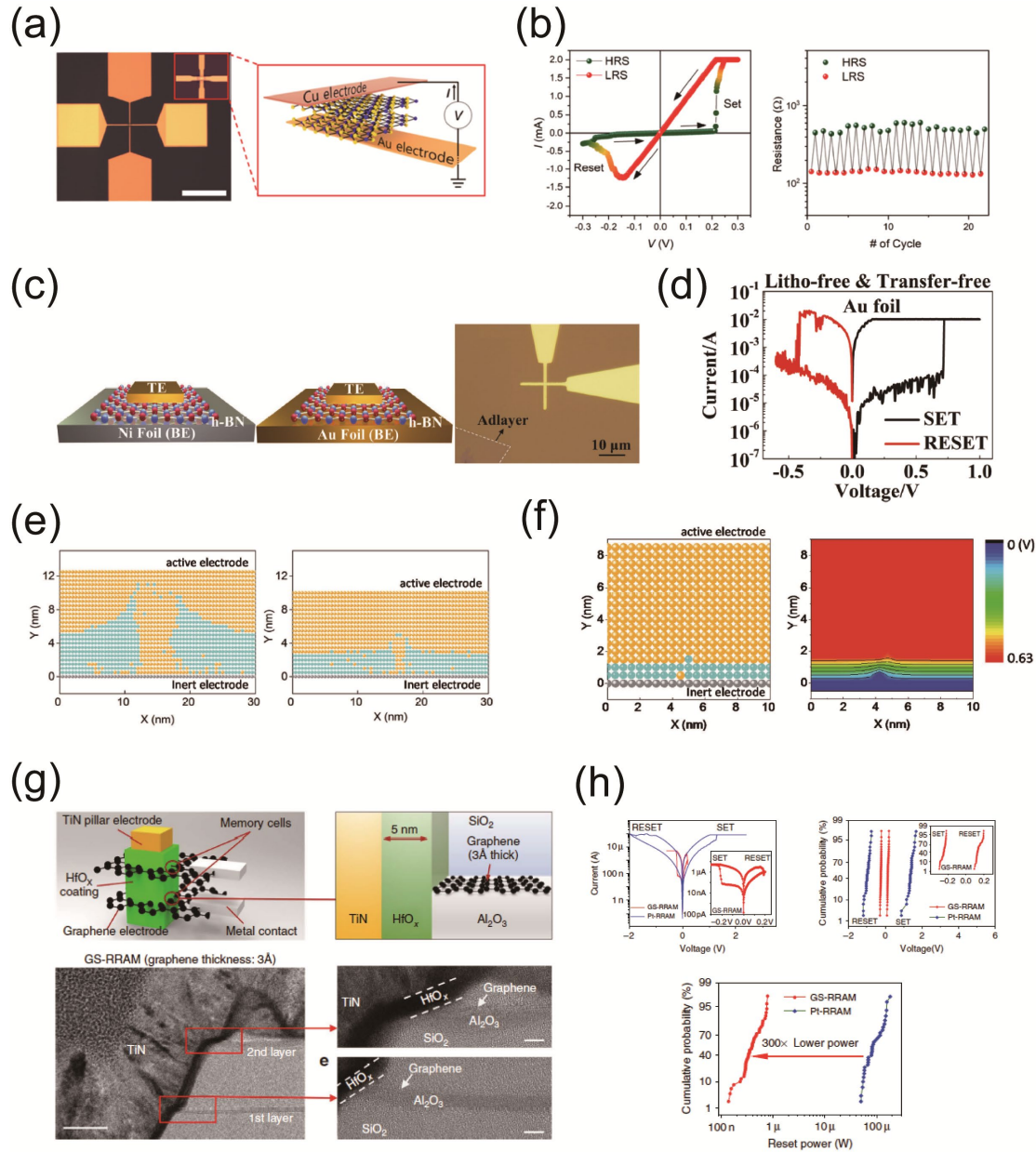

Figure 5. 2D layered materials for improving device integration density. (a). Optical image and illustration of bilayer h-BN memristive devices. (b). Left panel: switching curves with ultralow SET/RESET voltages; right panel: endurance test. (c). Schematic of memristive devices with monolayer h-BN as active layer. (d). Switching curve of device. (e). Simulations of filament formation with different h-BNOx thickness of 4.5 nm (left panel) and 2.25 nm (right panel). (f). Left panel: simulation of filament formation in 0.9 nm thick h-BNOx; right panel: electrical potential profile during filaments forming in the same film. (g). Illustration of vertical graphene-based memristive devices and high-resolution TEM images. (h). Improved performance in SET/RESET voltages and RESET power consumption compared to Pt-based counterparts. (a)-(b), reproduced with permission[112]. Copyright 2019, American Chemical Society Publishing. (c)-(d), reproduced with permission[113]. Copyright 2019, John Wiley & Sons, Inc. (e)-(f), reproduced with permission[80]. Copyright 2017, John Wiley & Sons, Inc. (g)-(h), reproduced with permission[111]. Copyright 2014, Springer Nature Publishing AG.

High-density integration of memristive devices is of crucial importance to the practical applications. Integration density is determined by lateral size of memristive devices. The limit for minimum lateral size is determined although integration has been demonstrated with 6-nm half-pitch and 2-nm critical dimension[114]. Theoretically, the cross-sectional area of filaments partially accounts for the minimum area of devices, as filaments laterally expand during penetrating through the active layer of memristive devices. The lateral expansion of conductive filaments would be restricted in the thin active layer, which is also limited by small operating current. Nevertheless, the thickness of active layer in most TMOs-based memristive devices ranges from 3 nm to 30 nm[3, 5], further reduction in thickness with controllable quality is challenging for thin-film deposition technologies[115, 116]. 2D layer-structured materials like transition metal dichalcogenides (TMDs) and hexagonal boron nitride (h-BN) are atomically thin and flat[36]. Use of 2D materials as active layer of memristive devices may push the thickness to atomic limit and reduce the SET/RESET voltages of memristive devices. As shown in Figure 5(a)-(b), Cu/bi-layer $MoS_2$/Au memristive devices show low SET (0.25 V) and RESET (-0.15 V) voltages[112], while the filament formation might be attributed to $Cu^{2+}$ diffusion through $MoS_2$. However, the low resistance (100 ~ 1000 Ω) for both LRS and HRS leads to a high-power consumption. This issue could be surmounted by the use of insulating h-BN[117] as active layer. The grain boundary in the h-BN active layer may serve as leaky paths to prevent dielectric breakdown in lateral direction[118]. By pushing the thickness of active layer to atomic limit, Ge et al. reported the first atomristor based on monolayer TMDs (i.e. $MoS_2$, $MoSe_2$, $WS_2$, $WSe_2$)[109]. Atomristors display unipolar and bipolar switching behaviors, and the operating current of atomristors could be reduced by replacing bottom metal electrode with graphene. Compared to TMDs, monolayer h-BN has a larger band gap, which may be more suitable for switching layer. By using CVD-synthesized monolayer h-BN, Wu et al. fabricated memristive devices with the thinnest active layer (~0.33 nm) and observed similar switching performance [113] with forming free feature (Figure 5(c)-(d)). This work presents a record thickness for switching layer of memristive device. Simulations have been performed to reveal that switching mechanism in monolayer h-BN might be due to substitution of metal ions into h-BN vacancies. Besides, Zhao et al. studied atomic filaments in memristive devices with ultrathin (0.9 nm) oxidized h-BN as active layer[80]. It is in general assumed that the formation and depletion of gap between the filament and the electrode are responsible for resistive switching. However, theoretical analysis suggests that the size limitation in this type of gap is around 2~3 nm[119], much larger than the thickness of h-BNO$_x$ being used. Therefore, the stable resistive switching in ultrathin h-BNO$_x$ indicates the existence of atomic filaments. To prove this point, simulations in filaments morphology with different h-BNO$_x$ thicknesses ($t_{ox}$) were performed, with results presented in Figure 5 (e). When $t_{ox}$ is relatively large, the filaments expand laterally within the size of several nanometers. Reducing $t_{ox}$ down to 0.9 nm (Figure 5(f)) gives rise to a single atomic chain filament. In this case, both SET and RESET voltages are larger than that of bilayer $MoS_2$ based memristive devices.

Three-dimensional (3D) integration offers a promising way for achieving high-density integration of memristive devices[120-123]. Lee et al. realized vertical integration of TiN/HfO₂/Gr memristive devices (Figure 5(g))[111] through fabricating edge contact between graphene and the active layer. As illustrated in left panel of Figure 5(g), oxygen vacancies in active layer are accumulated near graphene edge, since the activation energy of $O^{2-}$ diffusion in graphene electrode (0.15-0.8 eV) is much lower than that of TiN electrode (0.98-2.1 eV). Different from $O^{2-}$ drift in TiN/HfO₂/Pt memristive devices, using the atomically sharp edge of graphene as electrode could generate a larger electric field, to make $O^{2-}$ more easily driven from active layer to graphene electrode. This assumption has been justified by the results of spatially resolved Raman spectrum. Memristive devices with graphene edge contact also display a low operating voltage of $\pm 0.2$ V and a dramatic reduction in RESET power (Figure 5(h)). In addition to graphene, the 2D layered materials family also includes semiconductors and insulators. By using 2D semiconductor as transistor channel material and 2D insulator as memristive material, respectively, researchers have experimentally demonstrated 3D monolithically integrated one-transistor one-resistor (1T1R) cells [124], and shown potential of 2D materials in future high-density integration of memristive devices.

Apart from improving the performance of vertical memristive devices, the unique physical properties of 2D layered materials can also be exploited to develop planar memristive devices with gate tunable functionalities. Fermi level in bilayer graphene is gate-tunable. Based on this property, gate-tunable SET voltage is achievable in the bilayer graphene based planar memristive devices[125]. Usually, grain boundaries which exist in the chemical vapor deposition (CVD) synthesized TMDs are undesirable for the development of high-performance electronic applications. Interestingly, the grain boundaries can be modulated to migrate under the action of an electric field. By exploiting this behavior, Sangwan et al. demonstrated gate-tunable resistance switching[126] in the MoS₂ planar transistors. The movement of intersecting- or bisecting-grain boundaries in planar MoS₂ has been proposed to account for the resistive switching. This proposal is corroborated by electrostatic force microscopy (EFM) images and spatially resolved photoluminescence (PL) results. Atomic chain formation and breaking may occur in the atomic junctions under an electric field. The properties have been used to demonstrate graphene-based planar memristive devices with excellent endurance and retention properties[127]. Besides, 2D layered materials are free of dangling bond at the surface. This property makes them fully compatible with the state-of-art planar silicon-based technology. Integrating 2D materials with silicon substrate may bring new functionality[128], and open up new opportunities for multi-functional memristive devices based on mixed dimensional materials[43].

At the end of this section, the role of different ions migration in the 2D materials based memristive devices is compared. Similar to the TMOs-based memristive devices, other ions migration, e.g. $B^{3+}$, $N^{3-}$ and $O^{2-}$, in 2D layered materials have been proposed to account for the growth and rupture of filaments in vertical memristive devices[57, 60, 80, 109, 112, 113, 129]. It is well known that ions migration in conventional memristive devices causes the formation of filaments, while electric-field-induced $S^{2-}$ migration in MoS₂ based planar structure memristive devices leads to the movement of grain boundaries and gives rise to resistive switching. Apart from intrinsic ion migration, migration of external ions (*e.g.* $H^+$ and $Li^+$) in 2D materials could also result in resistive change, which would be discussed in detail in next section.[130-132] Finally, to make a comparison among key parameters, different types of 2D materials based memristive devices are listed (Table 1).

Table 1. Key parameters of 2D materials based memristive devices

| Material type | Thickness (nm) | TE/BE | Polarity | $V_{SET}$/ $V_{RESET}$ (V) | Operation current (A) | Endurance (cycles) | ON/OFF ratio | Retention (s) | Refs |
|---|---|---|---|---|---|---|---|---|---|
| MoTe$_2$ | 12 | Ti/Au | bipolar | 2.9/-1.75 | $10^{-8}$~$10^{-7}$ | NA | $10^5$~$10^6$ | $10^3$ | [63] |
| h-BNO$_x$ | 0.9 | Ag/Gr | bipolar | 0.65/-0.5 | $10^{-10}$ | 100 | 100 | $10^3$ | [80] |
| (PEA)$_2$PbBr$_4$ | 40 | Gr/Au | bipolar | 3/-1 | $10^{-11}$ | 100 | 100 | $10^3$ | [86] |
| MoS$_{2-x}$O$_x$ | 20-40 | Gr/Gr | bipolar | 1.2/-1.5 | $10^{-3}$ | $10^7$ | 100 | $10^5$ | [57] |
| MoS$_2$ | ~1.3 | Cu/Au | bipolar | 0.25/0.2 | $10^{-3}$ | 20 | 4 | $10^4$ | [112] |
| MoS$_2$ | ~10000 | Ti/Ti | bipolar | 80/-80 | $10^{-4}$ | 500 | 100 | $10^6$ | [133] |
| h-BN | 0.33 | Au/Au | bipolar | 3/-1.5 | $10^{-2}$ | 50 | $10^7$ | $10^5$ | [113] |
| | | | unipolar | 2/1 | $10^{-3}$~$10^{-2}$ | NA | NA | NA | |
| MoS$_2$ | ~0.6 | Au/Au | bipolar | 1.0/-1.25 | 0.01-0.1 | 150 | $10^4$ | $10^5$ | [109] |
| | | | unipolar | 3.5/1 | $10^{-4}$~$10^{-2}$ | NA | NA | NA | |
| | | Gr/Gr | bipolar | 4/-5 | $10^{-4}$ | NA | NA | NA | |
| MoSe$_2$ | ~0.6 | Au/Au | bipolar | 3/-1 | $10^{-4}$~$10^{-2}$ | NA | NA | NA | |
| | | | unipolar | 3.5/0.8 | $10^{-4}$~$10^{-2}$ | NA | NA | NA | |
| WSe$_2$ | ~0.6 | Au/Au | bipolar | 3/-1 | $10^{-3}$~$10^{-2}$ | NA | NA | NA | |
| | | | unipolar | 3.5/1.5 | $10^{-3}$~$10^{-2}$ | NA | NA | NA | |
| WS$_2$ | ~0.6 | Au/Au | bipolar | 2/-0.75 | $10^{-3}$~$10^{-2}$ | NA | NA | NA | |
| | | | unipolar | 1.5/1.5 | $10^{-3}$~$10^{-2}$ | AN | NA | NA | |
| h-BN | ~5 | Ti/Au | bipolar | 2/-1 | $10^{-2}$ | >100 | NA | NA | [60] |

## 3. Applications in neuromorphic computing

Neuromorphic computing is promising in solving the challenge that the data intensive tasks are inefficiently processed in computers with Von Neumann architecture. Some models of neural cells and systems can be implemented by various types of neuromorphic devices, via mimicking behaviors of neural cells and architectures of neural system. In these neural models, learning and memory processes depend on synaptic plasticity (also known as weight). The plasticity is considered as the ability of synapse to relay neural signals from one neuron to another, analogous to the definition of electrical resistance. The nonvolatile and continuous change of resistance of memristive devices can emulate the synaptic plasticity, thus artificial synaptic devices[15] can be used for implementation of neuromorphic computing. As shown in Figure 6, artificial synaptic devices are able to emulate basic synaptic activities like spike-timing-dependent plasticity (STDP), excitatory/inhibitory postsynaptic potential current (EPSC/IPSC), paired-pulse facilitation/depression (PPF/PPD) as well as short-/long-term plasticity (STP/LTP), which are building blocks of neuromorphic computing[3, 24, 134-138]. The emulation of transition between opposite biological behaviors, like excitatory/inhibitory and facilitation/depression, requires fabrication of synaptic devices with high tunability. However, electrical behaviors of synaptic devices with sandwich structure are fully controlled by vertical electric field. Thus, the traditional TMOs based devices are unable to exhibit diverse electrical characteristics, leading to a limitation in mimicking behaviors of bio-synapses. Different from TMOs-based devices, more degrees of freedom are available to be controlled in 2D materials, enabling us to utilize either electronic or optical properties to develop highly-tunable electronic synaptic devices or optoelectronic synaptic devices.

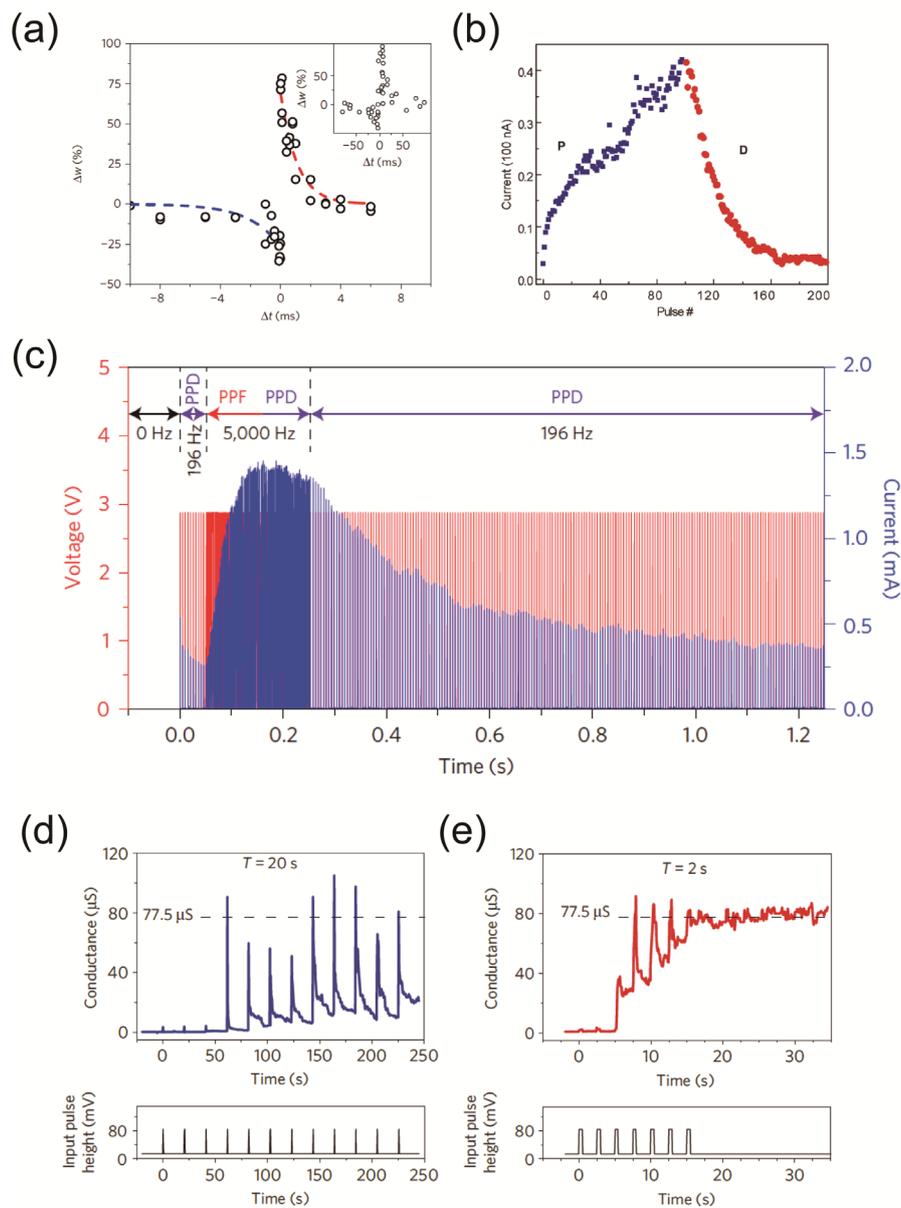

Figure 6. Mimicking bio-synaptic behaviors with artificial synaptic devices. (a). Spike time dependent plasticity. (b). Excitatory (blue points) and inhibitory (red points) postsynaptic potential current (EPSC and IPSC). (c) Paired-pulse facilitation and paired-pulse depression (PPF and PPD). (d)-(e). Short-term and long-term plasticity (STP and LTP). (a), reproduced with permission[59]. Copyright 2017, Springer

## 3.1. Electronic synaptic devices

2D layer-structured materials possess many unique physical properties, such as distinct lattice phases (e.g. 2H, 1T' and $T_d$), gate-tunability and so on. The phase transition and field effect offer new working mechanisms for designing novel synaptic devices. Based on distinct operating principles of devices, planar and vertical electronic synaptic devices can be constructed, which will be reviewed below.

### 3.1.1. Planar devices

Distinct lattice phases of 2D layered materials exhibit different physical properties. 2H phase corresponds to semiconducting transport properties while 1T' phase shows metallic behaviors in electrical transport. Exploring the lattice transitions among distinct phases provides new opportunities for developing novel synaptic devices.

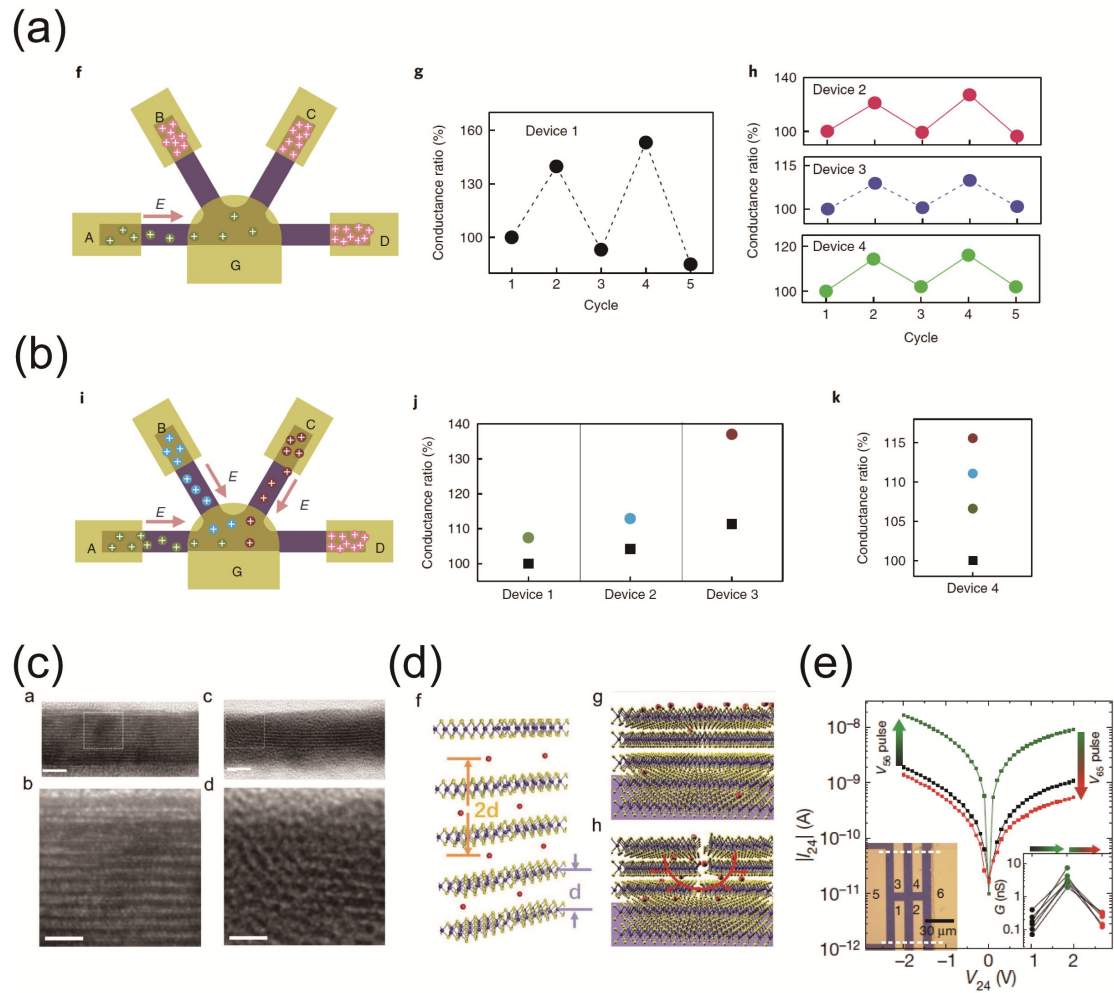

Figure 7. 2D materials based planar electronic devices: Ionic modulation of $Li_xMoS_2$ synaptic devices for emulating synaptic competition (a) and cooperation (b). (c). The cross-section SEM images of $WSe_2$ flakes after 60 (left panel) and 11115 (right panel) pulses of gate voltage. (d). Illustration of lithium ions intercalation into $WSe_2$. (e). Electrical behaviors of memtransistors based on polycrystalline monolayer

Ions intercalation in 2D layer-structured materials gives rise to lattice phase transition and provides a powerful tool to control electrical transport properties of 2D materials. Zhu et al. developed two-terminal planar $MoS_2$ memristive devices by intercalating $Li^+$ ions into $MoS_2$ flake[140]. In-plane electric field controls the direction of $Li^+$ ions lateral migration and locally triggers the phase transition from 2H to 1T' in $MoS_2$ channel. This local transition causes the change of contact resistance and results in non-volatile resistance switching. Note that resistive change in the $MoS_2$ regions would redistribute the amount of $Li^+$ ions near common electrode G and result in corresponding changes in resistance of other channel regions. Inspired by the behaviors of electrical transport, they further fabricated multi-terminal devices to emulate synaptic cooperation (Figure 7(a)) and competition (Figure 7(b)).

Different from two-terminal ionic devices, three-terminal ionic gating devices based on 2D layered materials enable to decouple write operations from read operations. Based on electrical double layer (EDL) at the interface between 2D layered materials and ionic liquid, Jiang et al. developed a $MoS_2$ ionic gating transistor with one modulatory gate and multiple input gates to emulate the synaptic behaviors like PPF and temporal filter[141]. Reconfigurable logic gates (e.g. AND and OR) have been demonstrated by varying gate voltages. More complex applications such as spatiotemporal coordinate and orientation recognition[142] can be achieved through assembling multiple input gates into designed arrays. Note that EDL modulated electrical transport is volatile, thus not suitable for emulating long-term synaptic plasticity. Gate-tunable ionic intercalation shows good performance in non-volatile switching, since 2D layer-structured materials would retain most of ions after gate voltage is removed. Based on the ionic intercalation mechanism, Yang et al. used quasi-2D insulating molybdenum oxide ($α$-$MoO_3$) to fabricate ionic liquid controlled transistors[132] and successfully emulate PPF and EPSP/IPSP with different shapes of input gate voltage pulse. The similar artificial synaptic behaviors were also observed in $α$-$MoO_3$ devices with solid electrolyte[143] and graphene devices with liquid gate[144].

Although the ionic intercalation mechanisms have been proposed to explain the observed resistance change, the direct evidence has been lacking. With employing high resolution transmission electron microscope (HRTEM), Zhu et al. studied the ionic intercalation in synaptic devices based on TMDs and phosphorus trichalcogenides[130] to reveal the intercalating dynamics of ions. The lithium ionic liquid serves as the ionic source to provide $Li^+$ ions in channel. $Li^+$ injection into $WSe_2$ films gives rise to the current ($I_{ds}$) change of the devices, which has been verified by HRTEM characterizations (Figure 7(c), (d)). Applying a few pulses would not cause significant structure change of $WSe_2$, while $WSe_2$ lattice structure exhibits clear disorder after stimulus of ten thousand voltage pulses. The apparent change of lattice structure is attributed to Li$^+$ intercalation, which is justified by the appearance of additional diffraction spots in SAED pattern of WSe$_2$ after thousands of voltage pulses.

Similar to the electrical field induced migration of external ions in the conventional vertical memristive devices, the movement of intrinsic ions in 2D planar materials also gives rise to stable resistive switching[131, 145]. Sangwan et al. demonstrated gate-tunable multi-terminal memtransistor based on polycrystalline monolayer MoS$_2$ and explored the device application in neuromorphic computing[133]. When an electric field is applied, the Sulfur vacancies in polycrystalline monolayer MoS$_2$ migrate. The non-volatile resistance change was observed with endurance of more than 400 circles and ON/OFF ratio of around 100. The CVD-grown MoS$_2$ is intrinsically n-type doped. Applying positive back gate voltage increases the electrical conductivity of MoS$_2$ channel, corresponding to lower resistances in the switching window. In six-terminal MoS$_2$ memtransistors, the channel resistance would change although the bias voltage is applied perpendicular to the channel direction. This electrical behavior is similar to the heterosynaptic plasticity in biological synapses (Figure 7(e)).

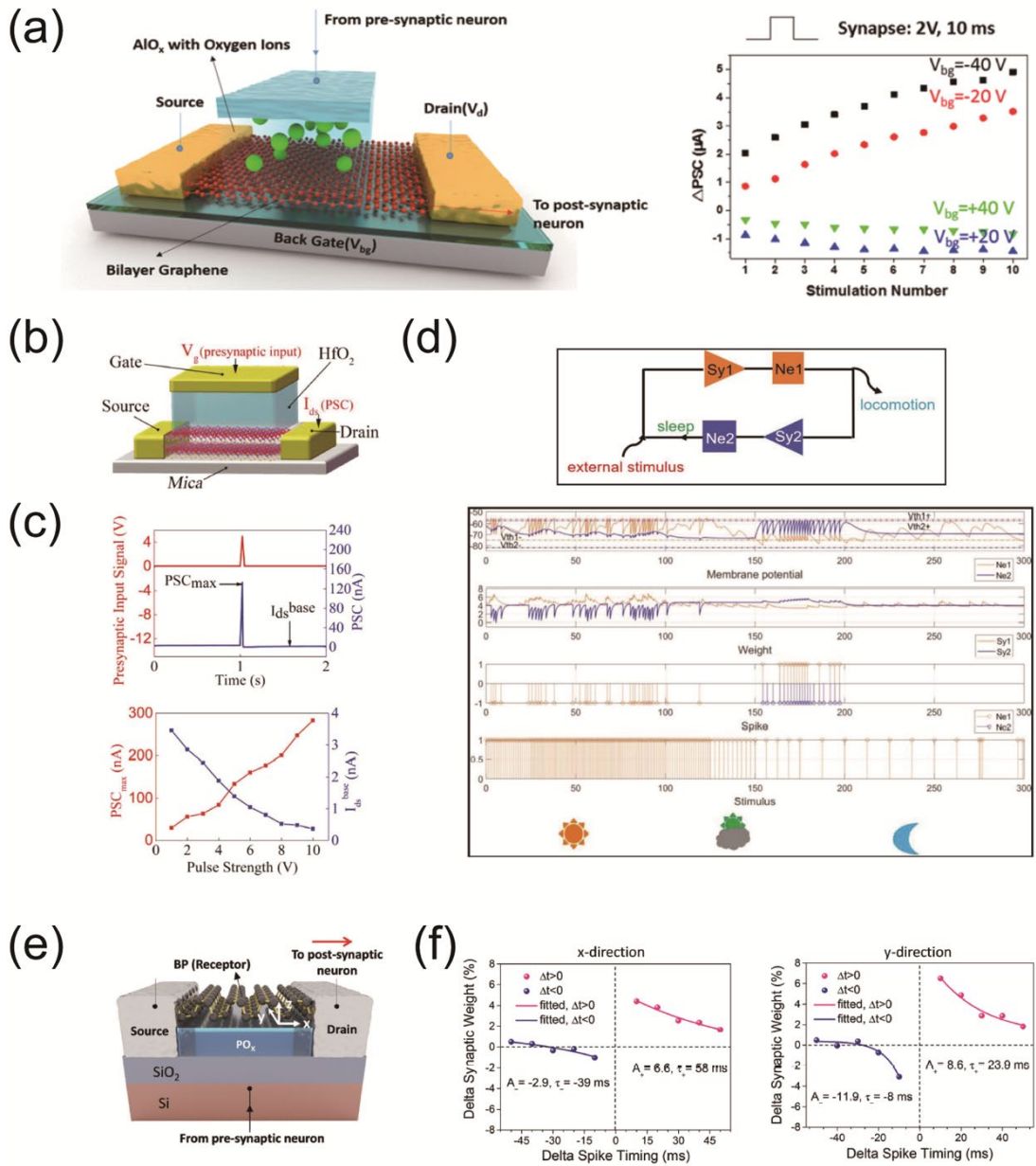

Figure 8. 2D layered materials based planar electronic devices: (a). Graphene synapse with gate-tunable plasticity. Left panel: the graphene-based synaptic device with AlO$_x$ dielectric layer; right panel: the polarity of $\Delta I_{ds}$ changes with back gate voltages. (b). Illustration of Bi$_2$O$_2$Se-based three-terminal memristive devices structure. (c). Increasing pulse strength shows independent tendencies of I$_{ds}$ peak and I$_{ds}$ base. (d). Heuristic recurrent neural circuit with two Bi$_2$O$_2$Se-based synaptic devices for "sleep-wake cycle autoregulation". (e). Schematic of black phosphorus transistor-like devices. (f). Difference of STDPs in x-direction and y-direction. (a), reproduced with permission[64]. Copyright 2015, American Chemical Society Publishing. (b)-(d), reproduced with permission[70]. Copyright 2019, John Wiley & Sons, Inc. (e)-(f), reproduced with permission[65]. Copyright 2016, John Wiley & Sons, Inc.

Apart from ionic migration induced resistance change of 2D layered materials, charge trapping at or near 2D materials has also been widely used for constructing artificial synaptic devices. As shown in Figure 8(a), dual-gate dynamic artificial synaptic devices based on twist bilayer graphene exhibit opposite electrical behaviors under different back gate voltages ($V_{bg}$) for the same stimulus from top gate ($V_{tg}$)[64]. This difference in the electrical response is due to $V_{bg}$ tunable doping-type transition in graphene. Applying a positive $V_{tg}$ pulse causes $O^{2-}$ migration from top dielectric layer ($AlO_x$) towards top gate electrode. The residual Oxygen vacancies with positive charges would trap electrons from graphene. The role of $V_{bg}$ in controlling device behaviors assembles that of neuromodulator (chemical messages to change activity of synapses) in synaptic plasticity transition between excitatory to inhibitory. By combing the effect of ionic migration and charge trapping together, a dual-mode artificial synapse with similar structure exhibits the behaviors of excitatory/inhibitory[67]. However, graphene based planar synaptic devices suffer from high power consumption due to the lack of band gap. By replacing graphene with $MoS_2$, the static power consumption of transistor-like artificial synapses could be reduced due to the suppressed leakage current in OFF state by applying gate voltage[146].

Using the charge trapping in dielectric layers enables to decouple short- and long-term plasticity. Zhang et al demonstrated a $Bi_2O_2Se$-based synaptic device (Figure 8(b)) [70] with top gate as presynaptic terminal. The peak current value of the device at stimulus moment is defined to emulate short-term plasticity ($PSC_{max}$), which depends on gate-tunable carrier density in $Bi_2O_2Se$ conducting channel, while the long-term plasticity ($I_{base}$) refers to the relaxed source-drain current level after stimulus (Figure 8(c)), which is determined by the amount of charges trapped in the dielectric film. Thus, changing carrier density in $Bi_2O_2Se$ channel and the amount of charge stored in the dielectric film separately leads to a concomitant and independently expressed short- and long-term plasticity. Based on orchestrated short- and long-term plasticity, they demonstrated "sleep-wake cycle autoregulation" in a heuristic recurrent neural circuit with two $Bi_2O_2Se$-based synaptic devices and two electronic neurons (Figure 8(d)). For the same $V_g$ stimulus, $PSC_{max}$ and $I_{base}$ show distinct variation tendency. Together with a recurrent circuit, electronic neuron 1 and 2 would fire alternately, which corresponds to the transition between sleep and wake states.

Individual artificial synaptic devices with isotropic material properties are able to emulate homogeneous behaviors of biological synapses. The competing behaviors among different bio-synapses, which are significant in neural activities, can be emulated in individual artificial synaptic devices based on 2D materials with low lattice symmetric properties. Black Phosphorus is very sensitive to air and its surface is easily oxidized to $PO_x$. $BP/PO_x$ based synaptic devices (Figure 8(e)) have been developed by using charge transferring and trapping effect between BP and $PO_x$. Based on anisotropic conductivity of BP and charge trapping in $BP/PO_x$ structure, heterogeneity of biological synapses were emulated[65]. As illustrated in Figure 8(f), the devices exhibit anisotropic STDP patterns with different amplitudes ($A_{+/-}$) and time constants ($\tau_{+/-}$) in x- and y-directions, respectively. The anisotropic signal transmission and weights modulation have been shown in multi-terminal BP synapse devices. Gaussian synapse can be emulated by connecting a p-type BP transistor with a n-type $MoS_2$ transistor in series, since the difference in transfer curve of BP and $MoS_2$ transistors under different gate voltages is close to a Gaussian distribution[147].

### 3.1.2. Vertical devices

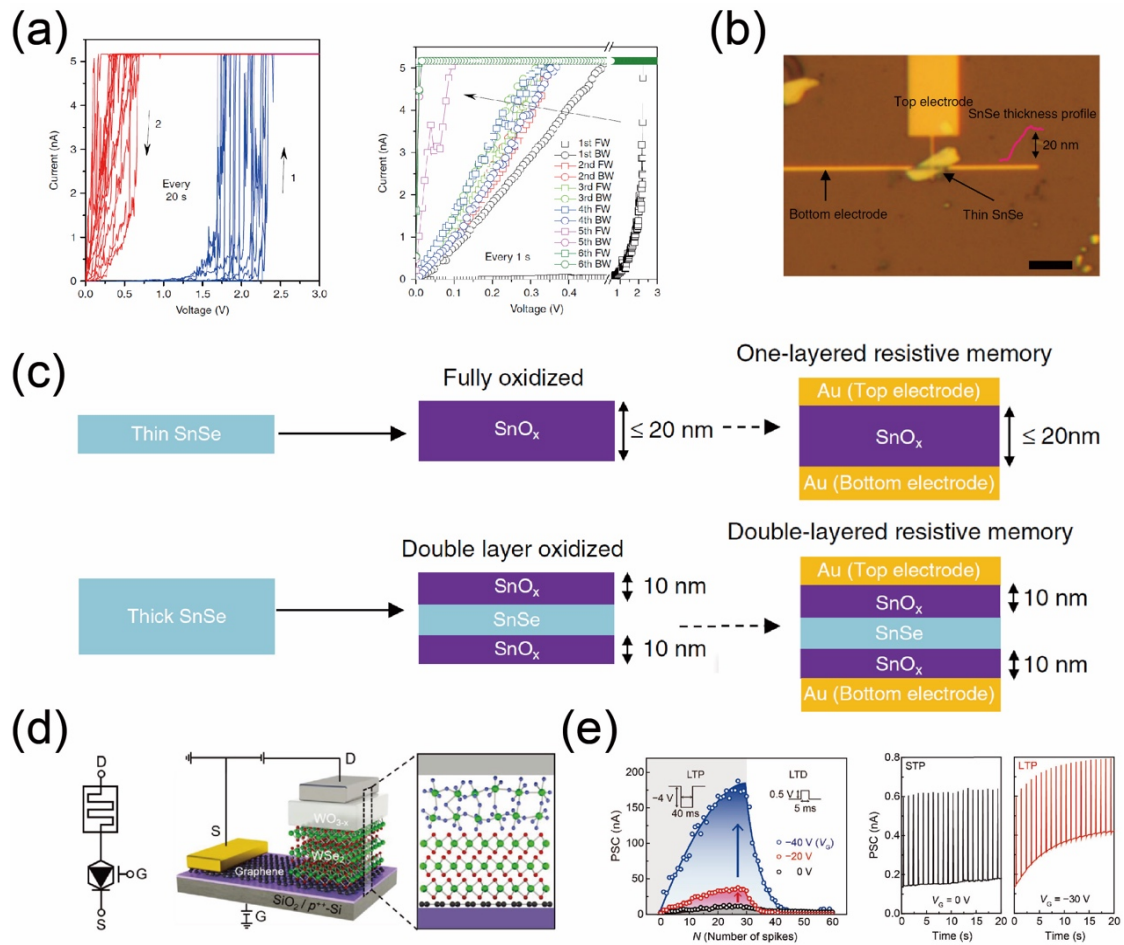

**Figure 9.** 2D layered materials based synaptic devices with vertical structure. (a). Electronic synapses based on multi-layer h-BN. Emulations of STP (left panel) and LTP (right panel) with different biases. (b). Optical image of SnSe device. (c). The oxidization process of SnSe flakes. (d). Illustration of synaptic barrister based on WO$_{3-x}$ memristive device and WSe$_2$/Gr barristor. (e). Changes of LTP, LTD, STP with different gate voltages. (a), reproduced with permission[60]. Copyright 2018, Springer Nature Publishing AG. (b)-(c), reproduced with permission[148]. Copyright 2018, Springer Nature Publishing AG. (d)-(e), reproduced with permission[129]. Copyright 2018, John Wiley & Sons, Inc.

Compared to planar devices, vertical devices are suitable for high-density integration although the coupled write and read operations[135] in memristive devices could limit their applications in mimicking diversity of biological neural activities. Different from memristive devices based on conventional oxides, 2D layered materials based vertical memristive devices possess additional degree of freedom for controlling and tuning, which is very useful in the emulation of diverse synaptic plasticity. In Pt/h-BN(~5 nm)/Cu based vertical artificial synapses, conductive filaments are only formed near the native defects in the active layer h-BN. Layered structure of h-BN effectively inhibits the lateral extension of conductive filaments, which controls the morphology of filaments[60]. When the interval of applied voltages decreases (Figure 9(a)), resistive switching changes from volatility to non-volatility, corresponding to a transition from STP to LTP. Similar behavior of restricting lateral growth of filaments has been also reported in Au/(PEA)$_2$PbBr$_4$/Gr vertical structure[86]. The use of wide bandgap (PEA)$_2$PbBr$_4$ (40 nm) reduces the operating current down to ~ pA. At this current level, basic behaviors of synapse plasticity including EPSC, IPSC, STP and LTP have been demonstrated. Cascading multiple memristive devices together though sharing electrodes allows for the independent formation and breaking of filaments in different active layers of different devices. This would give rise to multiple resistive states and interesting applications. This idea has been exemplified by the fabrication of vertical Au/SnO$_x$(~10 nm)/SnSe(20~180 nm)/SnO$_x$(~10 nm)/Au as shown in Figure 9(b)-(c), which could be applied in implementing algorithm of Markov chain and generating random numbers[148]. Recently, this idea has been further extended to 2D layered materials. By sandwiching graphene with two h-BN (~15 nm) flakes, Sun et al demonstrated separate control of formations and ruptures of filaments in two h-BN flakes [149] to achieve self-selective memory, which is promising in minimizing sneak current in large-scale memory operation.

Apart from tuning the resistive states in the cascaded memristive devices in an indirect way, gate-tunable interfacial properties of vdW heterostructures provide an alternative approach to tune resistive states by influencing formation and rupture of the filaments. In traditional TMOs-based synaptic devices, modulating the shape of LTP/LTD and realizing the transition between STP and LTP usually rely on the variation of the amplitudes and frequency of input voltage pulses[60, 86]. However, it is different in Ag/WO$_{3-x}$/WSe$_2$/Gr vertical devices. The modulation and transition could be realized by applying gate voltages, which dramatically simplifies the control of input signals[129]. The device was fabricated by cascading WO$_{3-x}$ based memristive device with WSe$_2$/Gr based barristor (Figure 9(d)). The vertical resistance of the barristor is gate-tunable as Schottky barrier height changes with the gate voltage[81]. For given bias across the vertical device, different gate voltages lead to different voltage drops on the WO$_{3-x}$ (~10 nm) memristive device. As shown in Figure 9(e), the electrical conductance of the barrister can be increased by applying negative gate voltage[81]. Consequently, a large voltage would be dropped across the memristive device to realize non-volatile switching. Without applying the gate voltage, the Schottky barrier is very high at the interface of WSe$_2$/Gr based barrister, thus source-drain voltage is most dropped at the interface. As a result, only a small voltage is divided onto WO$_{3-x}$ based memristive devices for switching. Realizing STP and LTP at a same device was also demonstrated in vertical heterostructures of h-BN/Gr/h-BN/MoS$_2$/Al$_2$O$_3$ [150]. Charge transfer occurs between different layers under the action of top and bottom gates. This charge transfer lead to resistive change of MoS$_2$, respectively. This transition in volatile and non-volatile resistive change could be used to mimic decision‐making action and in ‐situ storage. Similarly, gate-tunable doping-type variation in the vertical heterostructures could be used to enhance the functionality and diversity of neuromorphic electronic devices. The voltage-tunable variation of band profiles at SnSe/BP junction leads to a rectification inversion. Together with the change in the polarity of source-drain voltage, reconfigurable STDP patterns were achieved[151] with excitatory and inhibitory response modes.

## 3.2. Optoelectronic synaptic devices

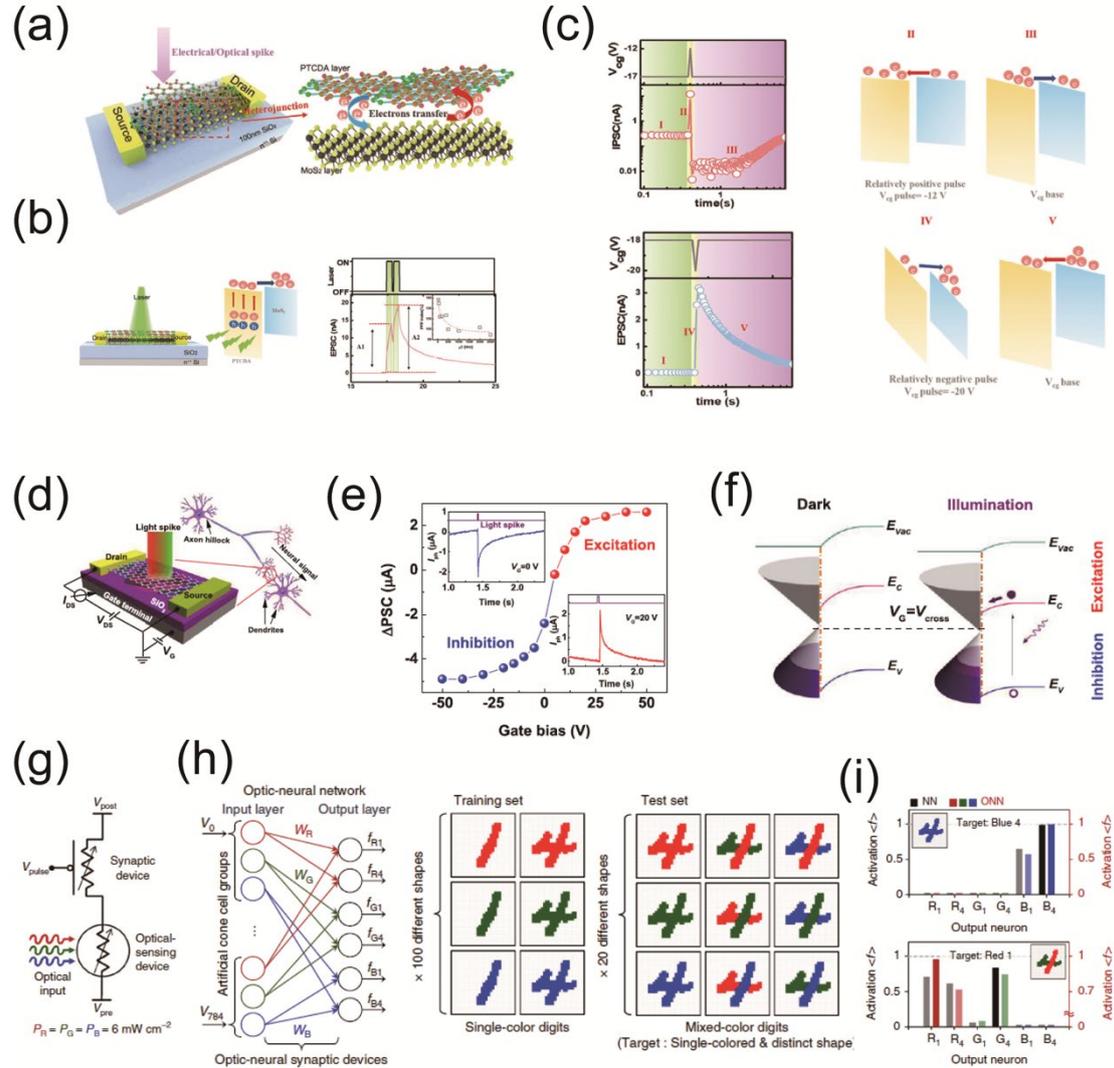

Figure 10. Optoelectronic synaptic devices based on 2D layered materials. (a). Illustration of MoS₂/PTCDA device and charge transfer between MoS₂ layer and PTCDA layer. (b). electrons transferring from PTCDA to MoS₂ leads to increase of source-drain current. (c) electrons transferring reversibly takes place between PTCDA and MoS₂ dependent on the relative polarity of applied electric pulse. (d). Illustration of device structure. (e). $\Delta I_{ds}$ variation at different $V_g$. (f). Graphene and single-wall carbon nanotubes junction. (g). Illustration of artificial optic-neural synapse. (h). Optic-neural network composed of artificial optic-neural synapses (left panel), and dataset for training and test (right panel). (i). The recognition results of two images: blue number "4" and red number "1" hidden in green number "4". (a)-(c), reproduced with permission[69]. Copyright 2019, John Wiley & Sons, Inc. (d)-(f), reproduced with permission[66]. Copyright 2017, Institute of Physics Publishing. (g)-(i), reproduced with permission[68]. Copyright 2019, Springer Nature Publishing AG.

In the electronic synaptic devices, electrical stimulus is used as signal input. Similarly, light stimulus can also be used as signal input to tune and control behaviors of the optoelectronic synaptic devices through modifying their electrical conductance [152-154]. Various optoelectronic synaptic devices based on different 2D layered materials and architectures have been proposed to emulate bio-synaptic features[66, 68, 69, 155-157]. Using laser with same or different wavelength as pre- and post-synaptic spikes, STDP has been realized in a hybrid optoelectronic device comprised of silicon crystals and WSe$_2$, in which boron-doped silicon crystals response to broadband light stimulus. By combining the electrical stimulus with optical one, Wang et al demonstrated MoS$_2$/PTCDA hybrid heterojunction based synaptic transistors Figure 10(a)[69] with dual-model operation: optical mode or electrical mode. In operation of the optical mode, MoS$_2$/PTCDA hybrid heterojunction was directly exposed to atmosphere. Light stimuli (532 nm) leads to a charge transfer from PTCDA to MoS$_2$. Varying back-gate voltage from zero to negative values causes a transition from non-volatile resistance change to volatile one (Figure 10(b)). Due to the charge transfer occurring at the heterojunction interface, PPF ratio of MoS$_2$/PTCDA devices is enhanced over the values reported in the previous works[66, 130, 132, 158]. Different from the optical mode, dielectric layer and control gate were added on the top of MoS$_2$/PTCDA hybrid heterojunction in the operation of electrical mode (Figure 10(c)). Electron transfer between MoS$_2$ and PTCDA determines the magnitude of current variation in the devices. To further increase the number of operating modes, a triple-mode MoS$_2$ synaptic transistor has been fabricated by using ionic gate. In this triple-mode device, the coexistence of Hebbian and homeostatic synaptic metaplasticity can be achieved through synergistic operation of back gate and ionic gate as well as light irradiation[155].

Usually, it is challenging for optoelectronic synaptic devices to emulate depression plasticity of biological synapses since light stimulus usually increases the electrical conductance, which seriously limits practical applications of such devices. To address this issue, Qin et al. proposed an optoelectronic synaptic device based on positive and negative photoresponse of hybrid materials (Figure 10(d)-(f) [66]). The hybrid materials were fabricated by inserting single-wall carbon nanotube (SWCN) films between graphene and SiO$_2$ substrates. In the device, current variation after light stimuli shows transition from positive to negative values as the gate voltage changes. This transition corresponds to a switchable synaptic plasticity between facilitation and depression (Figure 10(e)). The physical mechanism for this transition is schematically demonstrated in Figure 10(f). Light illumination excites electrons in valence band of SWCN valence band into conduction band of graphene. Then the photogenerated electrons recombine with the holes in negatively gated graphene. As a result, the suppressed conductance of the device emulates the depression behavior. Under the positive gate voltage, the light illumination increases the conductance of devices and emulates the features of synaptic facilitation.

Optoelectronic synaptic devices also show promising applications in pattern recognition, which is one of the basic cognitive functions of human being. Different from other neuromorphic devices that concentrate on processing information from image sensors[25, 26, 28, 159-161], optoelectronic synaptic devices combine image sensing and information processing together to drastically improve efficiency of data processing[162]. Seo et al. have recently reported colorful patterns recognition with artificial optical-neural synapses by connecting one $WSe_2$/h-BN heterostructure device with one $WSe_2$/Weight control layer (WCL)/h-BN device (Figure 10(g)) [68] in series. In this hybrid structure, $WSe_2$/h-BN device serves as the role of photodetector and transduces the light signal to electrical signal to input into the connected device. $WSe_2$/WCL/h-BN device shows multiple electrically resistive states, by controlling the amount of charges trapped in WCL. Assembling the hybridized optoelectronic devices into arrays (left panel of Figure 10(h)) enables to form an optical-neural network. This artificial network shows capability of realizing colorful recognition of MNIST numerical patterns (Figure 10(i)).

## 4. Challenges and outlook

As reviewed above, it is very promising to exploit the unique properties of 2D layered materials to well address the issues that the TMOs-based memristive devices are facing. But the associated challenges cannot be overlooked. The first challenge is the incompatible fabrication. Based on foundry-compatible fabrication facility, previous works have demonstrated TMOs-based small-scale memristive crossbar arrays. Using 2D layered materials indeed helps improving the performance of TMOs-based memristive devices from many aspects. However, a challenge is posed to the fabrication of large-scale high-performance memristive crossbar arrays with 2D materials involved in each cross-point. Because either transferring such large-area 2D materials onto TMOs or direct growth of the large-area 2D materials on the TMOs goes beyond the capability of what has been developed so far in the 2D materials community. In the near future, other promising approaches *e.g.* atomic layer deposition, may be much exploited to integrate 2D materials on to memristive crossbar arrays[163, 164]. Besides, some prototype demonstrations of 2D layered materials applications in neuromorphic computing are achieved on the basis of mechanical approach. The challenge in the synthesis of large-area 2D layered materials also limits their practical applications in neuromorphic computing. Although major progresses in large-area synthesis of monolayer 2D materials such as graphene[165], TMDs[166] and h-BN[164] have been achieved, it remains challenging to grow large-area 2D layered materials with controlled thickness. Recently, a few interesting device concepts have been proposed[68, 129], but their promising applications are also limited due to the lack of transformative technology in the fabrication of large-area vdW heterostructures.

Most of memristive and synaptic devices rely on the trapping effect or defects in or near 2D layered materials, which is not suitable for developing robust and high-performance devices. Novel memristive devices may be built by exploiting other unique physical properties of 2D layered materials, such as electrically-tunable phase transition and gate-controlled spin dynamics. Phase change memory materials have been widely used in memory[167-169] and neuromorphic computing[17, 159]. In particular, Intel has recently used this technology in its commercialized 3D X-bar point storage products. This would inspire researchers among 2D community to explore distinct phase transition mechanisms of 2D layered materials for proof-of-concept applications. Recent work has demonstrated that phase transition from 2H-MoTe$_2$ to 2H$_d$ structure shows promising application in high-performance memristive devices[63]. Unlike the phase transition occurring in traditional bulk materials, phase transitions in 2D layered materials are gate-tunable [85, 170, 171]. With continued effort along this direction, practical applications in memory or neuromorphic computing may be achievable in the near future. In addition, spin offers a new degree of freedom for memristive devices designing. Previous studies have demonstrated attainable vowel recognition[172, 173] based on spin transfer torque nano-oscillators, which shows feasibility of combing spintronic devices and neuromorphic computing together. The feasibility is illustrated by the recent work using the current-driven spin-orbit torque reversal in heterostructures[174]. Electrically tunable 2D layered magnetic materials also provide an ideal platform for building a variety of memristive devices[175-179]. The exact device structure and operating principles of memristive devices are further diversified by vdW heterostructures comprised of 2D layered magnetic materials[179].

2D materials based memristive and neuromorphic applications have achieved significant progresses. However, most of previous studies focus on individual devices. Little attention has been paid onto the construction of functional circuits based on 2D materials devices, drastically hindering their applications in neuromorphic computing, where complex functionalities may be required. In fact, neural microcircuits made of few synapses and neurons are widely distributed in neural systems. Some of microcircuits have simple connection architectures and are able to process information in a simple way[180], which are important to cognitive recognition. By assembling several 2D materials based neuromorphic devices to artificial neural microcircuits, more complex functionalities like sound localization and pattern recognition are expected to be implemented in experiments[142, 181].

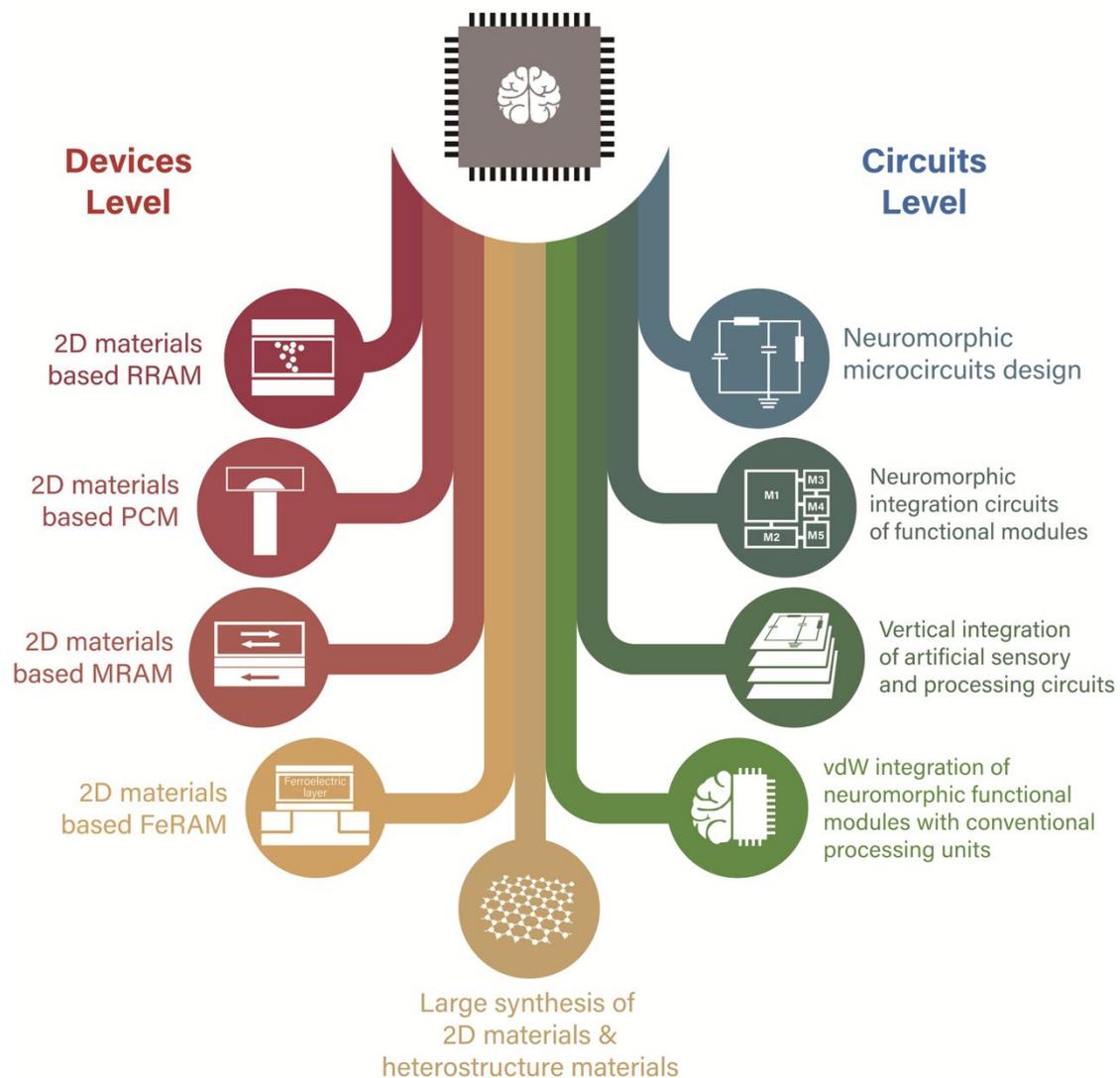

Figure 11. Roadmap of developing neuromorphic chips with integrated sensors, memory and processors based on 2D layered materials.

In summary, 2D layered materials show huge potential in development of memristive and neuromorphic devices with low power consumption and multifunctionalities. Future studies of memristive and neuromorphic devices based on 2D layered materials may proceed in different levels, as shown in Figure 11. In devices level, relentless search of new 2D materials and new vdW heterostructures will fuel the development of 2D materials based memristive and synaptic devices with new working mechanisms. Unique properties of 2D materials would help extending their applications in memory and neuromorphic devices, especially mimicking the behaviors of other biological cells like neurons[182]. In circuits level, advances in large-scale and high-quality growth of 2D materials as well as vdW heterostructures would be not only crucial to design more complex 2D materials based functional circuits but also beneficial to realize powerful neuromorphic computing based on integrated circuits. In chips level, new architectures of neuromorphic computing would integrate abilities of sensing, memorizing and processing by combing 2D materials based memristive devices, neuromorphic devices as well as sensors together. This approach may endow the neuromorphic chips with information processing ability of high-efficiency and low power consumption and offers a promising pathway for future technology revolution.

**Acknowledgements**

The authors C-Y. Wang, C. Wang and F. Meng made equal contribution to this work.

This work was supported in part by the National Key Basic Research Program of China (2015CB921600), the Fund for Creative Research Groups (61921005), the National Natural Science Foundation of China (61974176，61625402, 61574076), and the Collaborative Innovation Center of Advanced Microstructures and Natural Science Foundation of Jiangsu Province (BK20180330), Fundamental Research Funds for the Central Universities (020414380122, 020414380084).